\documentclass[prl,twocolumn,amsmath,amssymb,10pt,aps,longbibliography,superscriptaddress,citeautoscript,bibnotes,floatfix]{revtex4-2}

\usepackage{graphicx} 
\usepackage{dcolumn}
\usepackage{bm}
\usepackage{graphicx,color,xcolor}
\usepackage{caption}
\usepackage{physics}
\usepackage{comment}
\usepackage{ulem}
\graphicspath{ {./Figures/} }
\usepackage{textcomp}
\usepackage[utf8]{inputenc}
\usepackage[T1]{fontenc}
\usepackage{amsmath}
\usepackage{amssymb}
\usepackage{amsthm}
\usepackage{physics}
\usepackage{graphicx}
\usepackage{subfigure}
\usepackage{tabularx}
\usepackage{multirow}

\usepackage{xcolor}

\usepackage[colorlinks=true,linkcolor=blue,citecolor=blue]{hyperref}
\begin{document}

\title{Observation of 1/3 fractional quantum Hall physics in balanced large angle twisted bilayer graphene}

\author{Dohun Kim}
\thanks{These authors contributed equally.}
\affiliation{Department of Physics and Chemistry, Daegu Gyeongbuk Institute of Science and Technology (DGIST), Daegu 42988, Republic of Korea}

\author{Seyoung Jin}
\thanks{These authors contributed equally.}
\affiliation{Department of Physics, Pohang University of Science and Technology, Pohang, 37673, Republic of Korea}
\affiliation{Center for Artificial Low Dimensional Electronic Systems, Institute for Basic Science, Pohang 37673, Korea}

\author{Takashi Taniguchi}
\affiliation{Research Center for Materials Nanoarchitectonics, National Institute for Materials Science, 1-1 Namiki, Tsukuba 305-0044, Japan}

\author{Kenji Watanabe}
\affiliation{Research Center for Electronic and Optical Materials, National Institute for Materials Science, 1-1 Namiki, Tsukuba 305-0044, Japan}

\author{Jurgen H. Smet}
\affiliation{Max Planck Institute for Solid State Research, 70569 Stuttgart, Germany}

\author{Gil Young Cho}
\thanks{gilyoungcho@kaist.ac.kr}
\affiliation{Department of Physics, Korea Advanced Institute of Science and Technology, Daejeon 34141, Republic of Korea}
\affiliation{Center for Artificial Low Dimensional Electronic Systems, Institute for Basic Science, Pohang 37673, Korea}
\affiliation{Asia-Pacific Center for Theoretical Physics, Pohang, Gyeongbuk, 37673, Korea}

\author{Youngwook Kim}
\thanks{y.kim@dgist.ac.kr }
\affiliation{Department of Physics and Chemistry, Daegu Gyeongbuk Institute of Science and Technology (DGIST), Daegu 42988, Republic of Korea}

\date{\today} 

\begin{abstract}
\noindent {\bf Abstract}\\
\noindent Magnetotransport of conventional semiconductor based double layer systems with barrier suppressed interlayer tunneling has been a rewarding subject due to the emergence of an interlayer coherent state that behaves as an excitonic superfluid. Large angle twisted bilayer graphene offers unprecedented strong interlayer Coulomb interaction, since both layer thickness and layer spacing are of atomic scale and a barrier is no more needed as the twist induced momentum mismatch suppresses tunneling.  The extra valley degree of freedom also adds richness. Here we report the observation of fractional quantum Hall physics at 1/3 total filling for balanced layer population in this system. Monte Carlo simulations support that the ground state is also an excitonic superfluid but the excitons are composed of fractional rather than elementary charges. The observed phase transitions with an applied displacement field  at this and other fractional fillings are also addressed with simulations. They reveal ground states with different topology and symmetry properties.
\end{abstract}

\maketitle
\noindent {\bf Introduction}\\
\noindent Large angle twisted bilayer graphene offers the unique opportunity to address fractional quantum Hall (FQH) physics in a regime with very strong interlayer Coulomb interactions in view of the atomically thin interlayer spacing $d$  of $0.33\ {\rm nm}$. In the presence of a perpendicular magnetic field of strength $B$, the intralayer Coulomb interaction strength is set by the inverse of the magnetic length $l_B = \sqrt{h/eB}$. For typical field strengths accessible in the laboratory, this length scale is larger by an order of magnitude and the ratio of the interlayer to the intralayer Coulomb interaction strength, equal to $l_B/d$, can reach values as high as 20. This regime is inaccessible in conventional semiconductor based bilayer systems in which typically only $l_B/d \sim 1$ is achieved~\cite{Wiersma, Kellogg, Tutuc, Zhang_2016, Zhang_2020, Liu_2017, Li_2017, Liu_2019, Li_2019}. The finite quantum well thickness required to host a high quality two-dimensional electron gas and the need to separate both quantum wells with a thick enough barrier of a larger band gap semiconductor in order to suppress tunneling and observe uncluttered interlayer Coulomb interaction physics impose unsurmountable restrictions on the minimum interlayer spacing in conventional semiconductor heterostructures.

In large angle twisted bilayer graphene suppressed tunneling comes for free. The relative rotation of the two layers in real space also causes a displacement of the valley Dirac cones of the two monolayers by the same angle in reciprocal space. This generates a large momentum mismatch among the electronic states of the valleys belonging to the adjacent layers. This effectively suppresses interlayer tunneling \cite{YW_2013, YW, DH, SY, Leiwang} and  also prevents band hybridization across a large density range, leaving the valley and layer degrees of freedom essentially intact. This is in sharp contrast with the small twist angle bilayer counterpart, where the interlayer hybridization is significant at all densities~\cite{Cao_2018_Superconductivity, Cao_2018_Correlatedinsulator, Cory_MATBG, Efetov_MATBG, AFYoung_2020MATBG}. The exceptional density tunability of 2D layers and the ability to systematically vary the density imbalance among the two layers across a wide range through the application of a displacement field with the help of a top and back gate further boost the versatility of large angle twisted bilayer graphene as a platform to study incompressible ground states induced by interlayer Coulomb interations.

Here, we report on a systematic study of the FQH-states that emerge in this system and on transitions among FQH-states with different topological and symmetry properties observed as the displacement field is tuned. Unanticipated FQH physics at total filling 1/3 is observed for balanced population. Monte Carlo energy simulations of suitable trial wave functions are invoked to identify the origin of the FQH features and the transitions among them.

This study was performed on a total of four hBN encapsulated and dual-gated twisted bilayer graphene samples with twist angles of 9°, 18°, 8° and 11°, referred to as device D1 to D4. A schematic cross section of the layer sequence as well as optical microscope images of the devices are shown in Supplementary Fig. 1 of the Supplementary Information (SI). Graphite layers serve as top and bottom gates and enable independent density control of the top and bottom graphene sheets. 

The total charge carrier density can be obtained from the relationship $n_{\text{tot}} = eB\nu_{\text{tot}}/h = (C_TV_T + C_BV_B)/e$ with $V_T$ the top gate voltage, $V_B$ the bottom gate voltage, $C_T$ the top gate capacitance, $C_B$ the bottom gate capacitance and $\nu_{\text{tot}}$ the total filling factor. Using the values of $C_T$ and $C_B$ values, we calculate the displacement field  $\overline{D}/\varepsilon_{int}$ which considers the chemical potential correction and screening effect in graphene. Details on the effects of interlayer screening can be found in Supplementary Fig. 8 and Supplementary Note 3 of SI.

\begin{figure*}[!tbh]
\begin{center}
\includegraphics{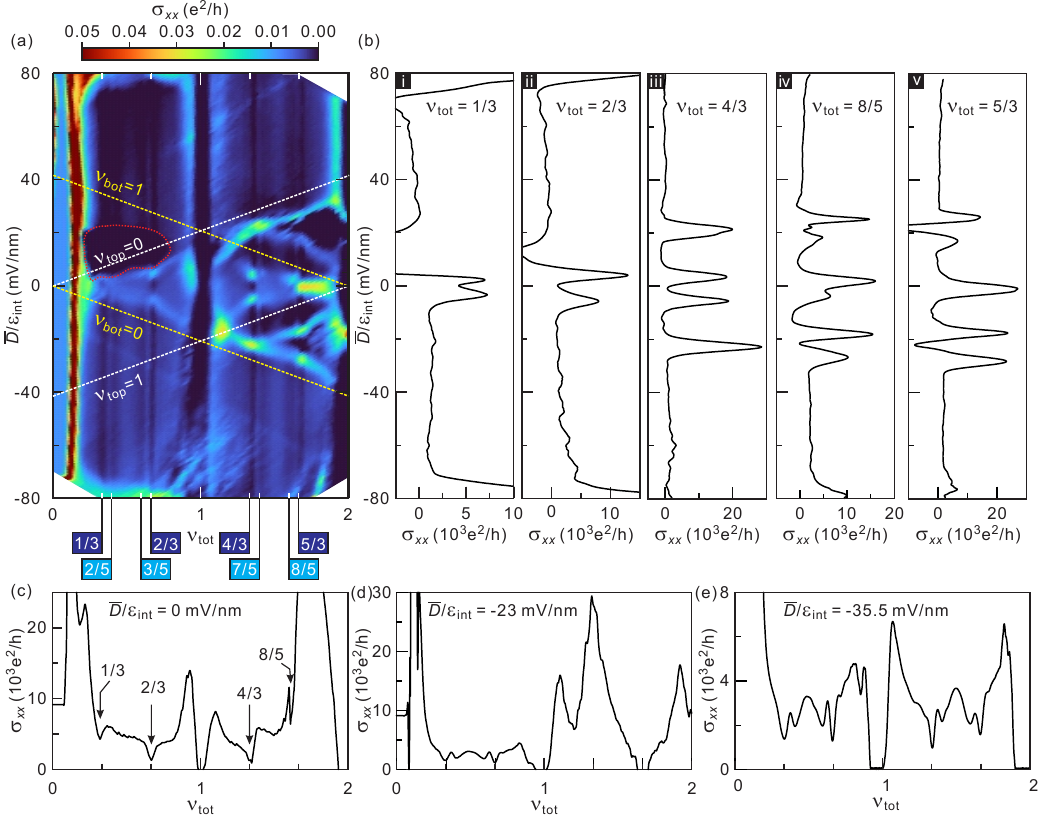}
\caption{{\bf The longitudinal conductivity $\sigma_{\rm xx}$ recorded on device D1 as a function of total filling $\nu_{\rm tot}$ and displacement field $\overline{D}/\varepsilon_{int}$ for a fixed magnetic field of 19 T and at a temperature of 30 mK.} {\bf (a)} Color rendition of the longitudinal conductivity in the ($\nu_{\rm tot}$, $\overline{D}/\varepsilon_{int}$)-plane. FQH-states related to two-flux composite fermions are marked at the bottom abscissa with colored boxes containing the total filling. Dotted lines highlight the conductivity minima caused by the condensation of the electrons in the top (white) or bottom (yellow) graphene layer into an integer quantum Hall state with filling 1 or 0. The region marked with red dotted line suffers from metal-graphene contact issue. {\bf (b)} Line traces of $\sigma_{\rm xx}$ as a function of the displacement electric field for fixed $\nu_{\rm tot}$ of 1/3, 2/3, 4/3, 8/5 and 5/3 in panels i through v from left to right. {\bf (c)} Line trace of $\sigma_{\rm xx}$ as a function of $\nu_{\rm tot}$ for a fixed displacement field of zero. {\bf (d)} Same as {\bf (c)}, but for $\overline{D}/\varepsilon_{int} = -23$ mV/nm. {\bf (e)} Same as {\bf (c)}, but for $\overline{D}v/\varepsilon_{int} = -35.5$ mV/nm.}
\label{Fig1}
\end{center}
\end{figure*}

\noindent{\large{\bf Results}}\\
\noindent {\bf Magnetotransport data}\\
Fig.~\ref{Fig1}(a) shows a color rendition of the longitudinal conductivity (\(\sigma_{xx}\)) for device D1 in the parameter plane spanned by \(\nu_{\text{tot}}\) and   $\overline{D}/\varepsilon_{int}$ for a fixed $B$-field of 19 T and a temperature $T$ of 30 mK. Several FQH-states appear as vertical dark blue features, each corresponding to a conductivity minimum at fixed total filling. They are marked by their fractional total filling on the abscissa. Line cuts for three different values of  $\overline{D}/\varepsilon_{int}$ are shown in panels c-e. Also at \(\nu_{\text{tot}} = 1\) an incompressible ground state is observed. This is consistent with previous studies \cite{YW, DH} that attributed this state to an interlayer coherent quantum Hall state caused by Bose Einstein condensation of excitons that form in the half filled lowest Landau levels of both layers. The yellow dotted lines in Fig.~\ref{Fig1}(a) running from $(\nu_{\text{tot}} = 0, \overline{D}/\varepsilon_{int} = 42)$ to $(2, 0)$ and from $(0, 0)$ to $(2, -42)$ highlight the location of the integer quantum Hall states for the bottom graphene sheet for filling $\nu_{\text{bot}} = 1$ and $0$. Similarly, white dashed lines in Fig. \ref{Fig1}(a) mark the location of these integer QH-states for the top graphene layer. The region with the shape of a diamond bounded by the $\nu_{\text{top(bot)}} = 0$ and $1$ lines contains a rich succession of maxima and minima in the conductivity. This is more clearly seen in Fig.~\ref{Fig1}(b) plotting vertical line cuts through this data set for some of the marked fractional total fillings.

For a balanced distribution of the charge carrier density among the two layers (i.e.~zero displacement field) and when interlayer Coulomb interactions are ignored, the bilayer system is anticipated to exhibit a conductivity minimum and condense in an FQH-state only for even numerator fractional total fillings equal to twice the filling of some prominent single layer FQH-state when both layers simultaneously condense in such a state. Here, however, that is not the case. A conductivity minimum also appears for balanced population when $\nu_{\rm tot}$ = $1/3$. The appearance of an FQH-state at this filling can only be understood as the result of strong interlayer Coulomb interactions. The minimum is separated from other minima at non-zero displacement field by conductivity maxima suggesting a transition to a FQH-state of different nature. Also for other fractional total fillings, numerous maxima separating adjacent conductivity minima appear in the vertical line cuts of Fig.~\ref{Fig1}(b) signaling transitions between FQH-states of different characters. Even though for a balanced density distribution fractional states with even numerator can in principle appear in the absence of interlayer Coulomb interactions, the actual ground states are likely also governed by the interlayer Coulomb interaction in view of its strength.  In order to assess the origin of the FQH-state at zero displacement field as well as the transitions at finite displacement we resort to Monte Carlo energy simulations~\cite{Morf, Jain_book} for the Hamiltonian describing large angle twisted bilayer graphene~\cite{Hunt, DH} and trial wave functions that are potentially suitable ground states causing the observed incompressible behavior at the fractional fillings in Fig.~\ref{Fig1}(a). We note that due to the additional valley degree of freedom, the number of plausible candidate wave functions to check for a given fractional filling can be rather large. The details of the model Hamiltonian and the Monte Carlo calculations as well as the procedure to construct trial wave functions are deferred to Supplementary Note $4 \mathrel{\sim} 7$ of the SI. Due to the imperfection of the metal-graphene contacts in our devices, \(\sigma_{xx}\) approaches zero in the range of 1/3 < \(\nu_{\text{tot}}\) < 2/3 and 5 < $\overline{D}/\varepsilon_{int}$ < 20 mV/nm, where the FQH-state is not observed and is indicated by a red dotted line.

\vspace{1em}
\noindent {\bf Theoretical phase diagram}\\
The simulation results are summarized in Fig.~2 (see also Supplementary Fig. 9 and Supplementary Note $9 \mathrel{\sim} 10$). In the left diagram (panel a), each vertical line represents a fractional filling for which a conductivity minimum is observed in experiment for a portion or the full range of the displacement field (Fig.~1). The color of each vertical line changes whenever a different trial wave function becomes energetically more favorable. The nature of each of these wave functions in panel a is elucidated in a square box demarcated by the same color in panel b for zero and positive displacement field. The top circles bounded by solid and dashed blue lines in each of these squares represents the valleys in the top layer centered around the $K_{\rm t}$ and $K'_{\rm t}$ symmetry points of the original Brillouin zone, respectively. Analogously, the bottom circles bounded by red solid and dashed lines refer to the two inequivalent valleys in the bottom layer. The filling of each circle coincides with the average occupation of that valley. The exact fractional filling can be determined with the help of the top legend. The total filled area within all four circles corresponds to the total filling. Each diagram visualizes the valley and layer symmetry breaking for that wave function. Two-component wavefunctions are depicted with a dumbbell shape (Supplementary Note 5 in the SI). The type of two-component state is encoded by the color of the dumbbell and included in the legend. More details about these wave functions as well as their topological properties can be found in Supplementary Table 3, Supplementary Table 4 and Supplementary Note $8 \mathrel{\sim} 10$  of the SI. For negative displacement fields the same wave functions are relevant but the roles of the top and bottom layer are simply reversed.

\begin{figure*}[!tbh]
\begin{center}
\includegraphics[width=\textwidth]{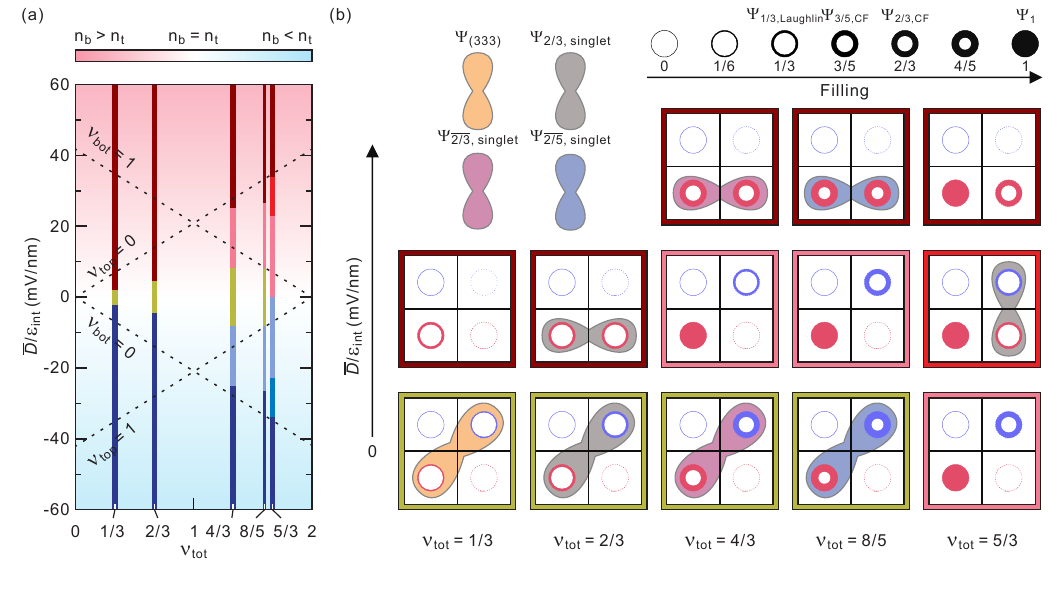}
\caption{{\bf Evolution of the FQH-states with displacement field} (a) Diagram summarizing the Monte Carlo energy simulations. Vertical lines at total fractional fillings change color depending on what wave function among the selected trial wave functions in Supplementary \sout{Note $4 \mathrel{\sim} 7$} Table 3 of the SI takes on the lowest energy. The properties of these lowest energy state are visualized in panel b. Yellow color is used for FQH-states satisfying the condition \(\nu_\text{top} = \nu_\text{bot}\). Red (blue) refers to partially layer-polarized states with \(\nu_\text{bot} - \nu_\text{top}=1(-1)\). Rest of the partially layer-polarized phases with \(\nu_\text{top} < \nu_\text{bot}\) (\(\nu_\text{top} > \nu_\text{bot}\)) are shown with light-red (light-blue) colors. Dark-red (dark-blue) is used for fully layer-polarized states with \(\nu_\text{tot}=\nu_\text{bot (top)}\).  The background color visualizes the degree of layer polarization of the twisted bilayer system. The four diagonal black dashed lines mark when  \(\nu_\text{bot}=1\), \(\nu_\text{top}=0\), \(\nu_\text{bot}=0\), and \(\nu_\text{top}=1\). (b) Schematics visualizing the occupation of the top and bottom layer and the two inequivalent valleys in each layer for the lowest energy trial wave function as the displacement field is tuned from zero to positive values. K and K´ valleys are shown by solid (K) and dashed (K´), blue (top layer) or red (bottom layer) lines. The degree of filling of each circle corresponds to the average occupation of that valley. The fractional filling can be determined with the help of the top right legend. The total filling factor relevant for each schematic is shown at the bottom and the color of the boundary surrounding a schematic is identical to the one used for that displacement in panel a. The colored dumbbells help to identify what two component state has been used to construct the wave function (top left legend). For completely or partially filled circles that are not part of a dumbbell, the wave function contains a one-component state such as the 1/3 Laughlin state for a valley of 1/3 filling or one filled state for a completely filled valley.}
\label{Fig2}
\end{center}
\end{figure*}

The blue and red color in the background of Fig.~\ref{Fig2}(a) visualizes the density imbalance, i.e. the degree of layer polarization. White corresponds to a balanced density distribution whereas red and blue color refer to preferred occupation of the bottom and the top layer, respectively. A comparison of the vertical lines with color changes at fractional fillings in Fig.~\ref{Fig2}(a) with the dark blue vertical lines with interruptions at the same fractional fillings in the magneto-transport data plotted in Fig.~\ref{Fig1}(a) reveals reasonable agreement. With the help of Fig.~\ref{Fig2} we can now discuss the most likely ground state for each fractional filling and how the ground state changes as the displacement field is varied.

As seen in  Fig.~\ref{Fig1}(b)-(i) a clear minimum in the conductivity is observed even at zero displacement field for $\nu_{\text{tot}} = 1/3$. The Monte Carlo energy simulations suggest that the underlying ground state is the interlayer coherent two component $(333)$-state (see Supplementary Fig. 10 and Supplementary Note 9, SI). This state can be considered the fractional analogue of the $(111)$-state occurring at total filling one due to the formation of excitons that share opposite charges between the two layers. Bose-Einstein condensation of the excitons into a superfluid gives rise to incompressible behavior. The same physical picture apparently applies here except that the excitons are formed out of quasi-particles with fractional charge 1/3 and -1/3. The topological properties, such as the ground state degeneracy on a torus, as well as the topological fractional excitations of this state are identical to those of the $1/3$ Laughlin state, but the wave function is spread equally and coherently across the two layers (see Supplementary Table 4). When the displacement field is increased, the interlayer coherent $(333)$-state disappears  and the fully layer-polarized, single component $1/3$-Laughlin state takes over (see Fig.~\ref{Fig2}(a) and top square in Fig.~\ref{Fig2}(b)-(i)). This is consistent with the displacement field driven transition in the experimental data shown in Fig.~\ref{Fig1}(a) and (b)-(i).

\vspace{1em}
\noindent{\bf{Magnetic field dependence of the $\nu_{\text{tot}} = 1/3$ FQH state}}\\

\begin{figure}
\begin{center}
\includegraphics{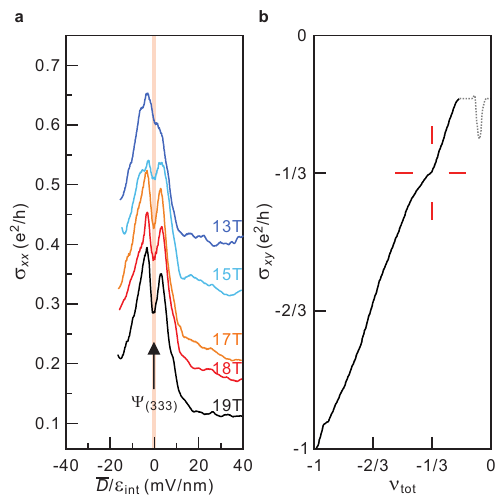}
\caption{{\bf Magnetic field dependence of $\sigma_{\rm xx}(\overline{D}/\varepsilon_{int})$  for $\nu_{\rm tot} = -1/3$ in device D2.} {\bf (a)} $\sigma_{\rm xx}$ as a function of $\overline{D}/\varepsilon_{int}$ for different magnetic field values. $\overline{D}/\varepsilon_{int} = 0$ is highlighted with an orange line. Here, the $(333)$ state develops. Curves are vertically offset for clarity. {\bf (b)} Hall conductivity $\sigma_{xy}$ as a function of $\nu_{\rm tot}$. Red lines emphasize the plateau for $\nu_{\rm tot} = -1/3$. The right most part of the curve is shown as a dotted line. The plateau is an artefact from the saturation of the lock-in amplifier as charge neutrality is approached and the Hall resistance rises rapidly.  All data were acquired at approximately 30 mK.}
\label{Fig3}
\end{center}
\end{figure}

The $\nu_{\text{tot}} = 1/3$ fractional quantum Hall (FQH) state at zero displacement field is observed not only in this sample but also in devices D2, D3, and D4 (Supplementary Fig. 2, Supplementary Fig. 4 and Supplementary Note 1, SI). Notably, transport data in these samples show enhanced performance under hole doping. Figure~\ref{Fig3}(a) highlights displacement field-dependent data recorded on device D2 at a fixed $\nu_{\text{tot}} = -1/3$ for varying magnetic fields. Panel (b) shows the Hall conductivity $\sigma_{xy}$ for a magnetic field of 19~T, calculated using device geometry and the tensor relation 

\[
\sigma_{xy} = \frac{R_{xy}}{(R_{xx}(w/l))^2 + R_{xy}^2} \times \frac{h}{e^2},
\]

where $w$ and $l$ represent the sample width and the distance between voltage probes. A quantized plateau is observed, while the minimum in $\sigma_{xx}$ appears from 15~T onwards, becoming more pronounced with increasing magnetic field. The $(333)$ state has not previously been reported for twisted bilayer graphene. Although a $\nu_{\text{tot}} = 1/3$ state was observed in double-layer graphene separated by a thin hBN layer \cite{Li_2019}, that study lacked theoretical support and used Corbino geometry, limiting quantized Hall measurement.

In contrast, the $\nu_{\text{tot}} = 2/3$ FQH state in double-layer systems is commonly observed without a displacement field, as it does not rely on interlayer Coulomb interactions. Instead, it can result from two decoupled $1/3$ Laughlin states in each layer when the interlayer distance $d$ greatly exceeds the magnetic length $l_B$, as extensively documented \cite{McDonald, Peterson, Geraedts, Jain_2020}. However, in regimes with strong interlayer interactions, other energetically favorable ground states may emerge. Trial wave functions considered include pseudospin singlet states \cite{Davenport, Jain_2020}, interlayer/intralayer Pfaffian states \cite{Wen_Pfaffian}, the particle-hole conjugate of the $1/3$ Laughlin state as a composite fermion (CF) state \cite{Jain_book, Davenport}, $\mathbb{Z}_4$ Read-Rezayi states \cite{Read-Rezayi, Wen_Z4}, and composites of two $1/3$ Laughlin states \cite{Jain_book, Jain_2020}. Monte Carlo simulations for $\nu_{\text{tot}} = 2/3$ revealed that the lowest-energy ground state at zero displacement field is an interlayer coherent pseudospin singlet state (bottom square in Fig.~\ref{Fig2}(b)-(ii)) with electrons occupying different valleys in the top and bottom layers. When displacement field increases, electrons shift entirely to the bottom layer, yielding $\nu_{\text{tot}} = \nu_{\text{bot}} = 2/3$ and $\nu_{\text{top}} = 0$, forming a fully layer-polarized pseudospin singlet state (top square in Fig.~\ref{Fig2}(b)-(ii)). This transition alters symmetry but preserves topological properties (see Supplementary Fig. 13 and Supplementary Note $8 \mathrel{\sim} 10$, SI).

\begin{figure}
\begin{center}
\includegraphics{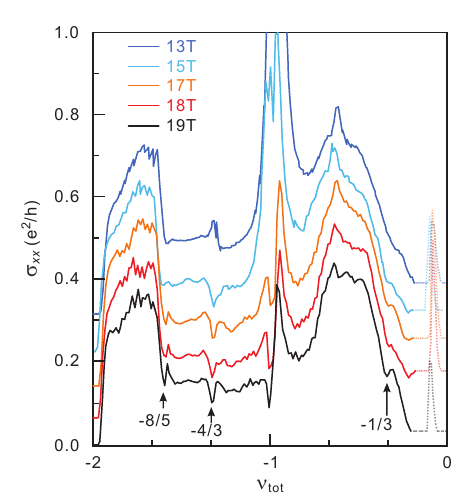}
\caption{{\bf Longitundinal conductivity in device D2 as a function of the total filling and for different fixed values of the magnetic fields.}  The FQH-states, $\nu_{\rm tot}$ = -1/3, -4/3, and -8/5, are highlighted. Curves are vertically offset for clarity. The flat minima around $\nu_{\rm tot}$ = 0 shown as faint dotted lines are artefacts caused by the saturation of the lock-in amplifier.}
\label{Fig4}
\end{center}
\end{figure}

Similarly, at $\nu_{\text{tot}} = 4/3$ and $8/5$, interlayer coherent pseudospin singlet states dominate without displacement fields, while large displacement fields induce fully layer-polarized states in one layer. Intermediate fields result in mixed states involving integer ($\nu = 1$) QH states in one layer and fractional ($\nu = 1/3$ or $3/5$) QH states in the other (Fig.~\ref{Fig2}(b)-(iii), (iv), Supplementary Fig. 14, and 16). Magnetic field-dependent data for $\nu_{\text{tot}} = -8/5$, $-4/3$, and $-1/3$ on device D2 are shown in Fig.~\ref{Fig4} (see also Supplementary Fig. 4). FQH states develop differently across cool-down cycles (see Section I, SI).

For $\nu_{\text{tot}} = 5/3$ (Fig.~\ref{Fig1}), the absence of FQH behavior at zero displacement field stems from challenges in constructing balanced-layer trial wave functions, such as two independent FQH states at $\nu_{\text{top}} + \nu_{\text{bot}} = 5/6 + 5/6$. Instead, partial layer polarization near zero displacement field combines an integer QH state in one layer with a 2/3 CF state in the other. Intermediate displacement fields produce a mix of interlayer coherent and integer QH states, transitioning to fully layer-polarized states at higher fields (Fig.~\ref{Fig2}(b)-(v), see also Supplementary Fig. 17 and Supplementary Note 10, SI).

\vspace{1em}
\noindent{\large{\bf Discussion}}\\
We conclude that large angle twisted bilayer graphene with strongly suppressed interlayer tunneling and a layer separation on the atomic scale is a powerful test bed for the exploration of correlation physics induced by interlayer Coulomb interactions of unprecedented strength. For balanced layer population interlayer coherent states appear such as (333)-state at  $\nu_{\text{tot}} = 1/3$, the fractional analogue of the superfluid exciton condensate observed in the integer quantum Hall regime.  
This state has been challenging to detect in double-layer systems based on conventional semiconductor heterostructures. However, it has previously been reported in a graphene/h-BN/graphene structure using Corbino geometry~\cite{Li_2019}, despite the absence of supporting theoretical calculations. In our study, we utilized Monte Carlo energy simulations with carefully chosen trial wave functions to confirm the ground state of not only the 1/3 state but also other FQH-states and their transitions across different filling factors.

\vspace{1em}  
\noindent{\large{\bf Methods}}\\
\noindent {\bf Device fabrication}

\noindent We used the dry pick-up technique  with an Elvacite stamp to assemble a graphite/h-BN/twisted bilayer graphene/h-BN/graphite/h-BN (from bottom to top) heterostructure, where the thickness of the h-BN was between 50 and 70 nm and the thickness of graphite was approximately 5 nm.\cite{Wang_2013, Kim_2019} Both graphites were used as the top and bottom gates. The heterostructure was then annealed at $600^{\circ}\text{C}$ in a forming gas (Ar/H$_2$) environment for 30 minutes. The Hall bar geometry and edge contact were defined using e-beam lithography and reactive ion etching with a mixture of CF$_4$ and O$_2$ and a power of 40 W. Finally, a metal electrode was deposited with Cr/Au (5/100 nm) by e-beam evaporator with a base pressure of $5 \times 10^{-7}$ torr \cite{DH, SY}.

\vspace{1em}  

\noindent {\bf Transport measurements.}

\noindent Transport measurements were performed with the help of low frequency lock-in techniques using an ac current of 100 nA and a frequency of 17.777 Hz in a dilution refrigerator with a base temperature of 30 mK. The samples were fabricated on top of a heavily doped silicon substrate with a 285 nm thick thermal oxide at the top. Apart from the top and back gate voltages applied to the graphite layers to tune the density and the displacement field, also a voltage of 70 V or -70 V (for electron or hole doping of the twisted graphene bilayer) was applied to the Si substrate in order to reduce the contact resistance in regions not covered by the graphite gate layers \cite{YW}.

\vspace{1em}  

\noindent {\bf Monte Carlo simulation.} 

\noindent We theoretically model the system of large angle twisted bilayer graphene by the Hamiltonian consisting of the Coulomb interaction term, the potential energy term due to the displacement field, the capacitive energy term due to the finite interlayer distance, and the phenomenological short-range interaction term (See Supplementary Note 4 of SI for details) \cite{DH,Hunt}. As briefly discussed in the main text, various states of different topology and symmetry are considered as candidate wave functions of the ground state. To determine the ground state, we employ Monte Carlo method ~\cite{Morf, Jain_book} for calculation of the energy for the candidate wave functions. For each state, the energy is calculated from \(10^6\) samples of particle configurations collected via Metropolis-Hastings algorithm. The algorithm is ahead optimized by acceptance rates and integrated autocorrelation times computed from the pre-run data. The acceptance ratio of the algorithm is the ratio of probability amplitudes. The detail is in Supplementary Note 6 of SI. 
\vspace{1em}

\noindent{\large{\bf Data availability}}\\
Source data for all main text figures are included in the Supplementary Information. Any other data supporting the findings of this study are available from the corresponding author upon request. Source data are provided with this paper

\noindent{\large{\bf Code availability}}\\
Key algorithms of theoretical calculations are fully clarified in the Supplementary Information.

\vspace{1em}

\noindent{\large{\bf References}}\\

\vspace{1em}

\noindent{\large{\bf Acknowledgement}}\\
We thank Klaus von Klitzing, Byungmin Kang, and Kwon Park for helpful discussions. Half of this research was supported by the Nano and Material Technology Development Program through the National Research Foundation of Korea (NRF) funded by Ministry of Science and ICT (No. RS-2024-00444725). S.J. and G.Y.C. are supported by the Samsung Science and Technology Foundation under Project Number SSTF-BA2002-05 and SSTF-BA2401-03, the NRF of Korea (Grant No.RS-2023-00208291, No.2023M3K5A1094810, No.2023M3K5A1094813, No.RS-2024-00410027) funded by the Korean Government (MSIT), the Air Force Office of Scientific Research under Award No.FA2386-22-1-4061, and the Institute of Basic Science under project code IBS-R014-D1. The work from DGIST was supported by the Basic Science Research Program NRF-2020R1C1C1006914, NRF-2022M3H3A1098408 through the National Research Foundation of Korea (NRF) and the BrainLink program funded by the Ministry of Science and ICT through the National Research Foundation of Korea (2022H1D3A3A01077468). We also acknowledge the partner group program of the Max Planck Society. J.H.S is grateful for financial support from the SPP 2244 of the DFG. K.W. and T.T. acknowledge support from the JSPS KAKENHI (Grant Numbers 21H05233 and 23H02052) and World Premier International Research Center Initiative (WPI), MEXT, Japan.

\vspace{1em}

\noindent{\large{\bf Author contributions}}\\
D. K, Y. K and G.Y.C conceived the project. D. K carried out the device fabrication and performed the low-temperature measurement with J. H. S and Y. K. The theory was performed by S. J and G. Y. C. The h-BN crystals were synthesized by T. T and K. W. All authors contributed to the manuscript writing.

\vspace{1em}

\noindent{\large{\bf Competing interests}}\\
The authors declare no competing interest

\end{document}